%% file: paper.tex
\documentclass[sigconf]{acmart}

\usepackage{soul}
\usepackage{xspace}
\usepackage{url}
\usepackage{graphicx}
\usepackage{listings,multicol} 
\usepackage[export]{adjustbox}
\usepackage{amsmath}
\usepackage{textcomp}
\usepackage{breakurl}
\usepackage{listings}
\usepackage{cleveref}
\usepackage{algorithm}
\usepackage{enumitem}
\usepackage[noend]{algpseudocode}
\usepackage{multirow}
\usepackage[most]{tcolorbox}
\usepackage{siunitx}
\usepackage{booktabs}
\usepackage{subcaption}
\usepackage{xcolor}

\newcommand{\ie}{\emph{i.e.,}\xspace}
\newcommand{\eg}{\emph{e.g.,}\xspace}

\newcommand{\toolname}{\emph{SuperParser}\xspace}
\newcommand{\javanorm}{\emph{Java-norm}\xspace}

\newcommand{\secpart}[1]{\subsection{#1}}

\definecolor{bg}{RGB}{255,249,227}

\title{Evaluating the Impact of Source Code Parsers on ML4SE Models}

\author{Ilya Utkin}
\email{utkin.ilya.nickolaevich@gmail.com}
\affiliation{%
  \institution{Saint Petersburg State University}
  \country{}
}
\author{Egor Spirin}
\email{spirin.egor@gmail.com}
\affiliation{%
  \institution{JetBrains Research}
  \country{}
}
\author{Egor Bogomolov}
\email{egor.bogomolov@jetbrains.com}
\affiliation{%
  \institution{JetBrains Research}
  \country{}
}
\author{Timofey Bryksin}
\email{timofey.bryksin@jetbrains.com}
\affiliation{%
  \institution{JetBrains Research}
  \country{}
}

\begin{document}

\begin{abstract}
As researchers and practitioners apply Machine Learning to increasingly more software engineering problems, the approaches they use become more sophisticated. A lot of modern approaches utilize internal code structure in the form of an abstract syntax tree (AST) or its extensions: path-based representation, complex graph combining AST with additional edges. Even though the process of extracting ASTs from code can be done with different parsers, the impact of choosing a parser on the final model quality remains unstudied. Moreover, researchers often omit the exact details of extracting particular code representations. 

In this work, we evaluate two models, namely Code2Seq and TreeLSTM, in the method name prediction task backed by eight different parsers for the Java language. To unify the process of data preparation with different parsers, we develop \toolname, a multi-language parser-agnostic library based on PathMiner. \toolname facilitates the end-to-end creation of datasets suitable for training and evaluation of ML models that work with structural information from source code. Our results demonstrate that trees built by different parsers vary in their structure and content. We then analyze how this diversity affects the models' quality and show that the quality gap between the most and least suitable parsers for both models turns out to be significant.
Finally, we discuss other features of the parsers that researchers and practitioners should take into account when selecting a parser along with the impact on the models' quality.

The code of \toolname is publicly available at \url{https://doi.org/10.5281/zenodo.6366591}. We also publish \javanorm, the dataset we use to evaluate the models:  \url{https://doi.org/10.5281/zenodo.6366599}.
\end{abstract}

\maketitle

\input{sections/01-introduction}
\input{sections/02-background}
\input{sections/03-astminer}
\input{sections/04-experiments}
\input{sections/05-results}
\input{sections/06-discussions}
\input{sections/07-conclusion}

\bibliographystyle{IEEEtran}
\bibliography{IEEEabrv,paper}

\end{document}

%% file: sections/01-introduction.tex
\section{Introduction}\label{sec:introduction}

The field of applying machine learning (ML) algorithms to source code rapidly grows~\cite{sharma2021survey}.
Earlier ML models represented code as a set of explicitly defined metrics or as plain text. The latter makes it possible to reuse approaches from the Natural Language Processing (NLP) domain~\cite{Ernst2017nlp_for_se}.
While any file with source code can indeed be seen as a text document, source code has a richer underlying structure compared to natural languages. Nowadays, there are increasingly more works that take this underlying code structure into account~\cite{alon2018code2seq, allamanis2020typilus, zugner2021codetrans}.

Every programming language has its own strict syntax that can be represented as a set of predefined rules determining all possible language constructions.
Parsing the raw source code with these rules allows representing it in the form of a parse tree, and then as an abstract syntax tree (AST).
By working with this structure and also with its possible extensions (\eg path-based representation~\cite{alon2018general}, graph-based representation~\cite{allamanis2017learning}) researchers improve results for various software engineering tasks: code summarization~\cite{alon2018code2seq}, type inference~\cite{allamanis2020typilus}, bug detection and repair~\cite{Hellendoorn2020Global}, malware detection~\cite{rusak2018ast}.

Aside from the ML domain, the representation of code as an AST is crucial for its analysis and execution: it is used by compilers, static analysis tools, and integrated development environments (IDEs).
As a result, there exist a lot of different parsers for building ASTs from source code. Most parsers are tailored for a specific problem and have their own features. For example, \textsc{Tree-sitter}~\cite{max_brunsfeld_2022_6326492} aims to work with incremental parsing, \textsc{ANTLR}~\cite{10.5555/2501720} makes it easier to create new languages by providing parser generators, and so on.
All the parsers have different use-cases and require different levels of information from code, therefore, they might build different ASTs.

Even though for most languages there are many available parsers, not all the research works that use information from ASTs to train ML models provide enough details on how they build syntax trees.
Machine learning models are very sensitive to the input data, and as different parsers build ASTs in a different manner, the choice of a parser might affect the model's quality. In turn, it might lead to unreproducible results or even incorrect model comparison if the models use different parsers.

In this work, we investigate whether the selection of a source code parser affects the final quality of ML models.
We studied the impact of the parser choice on two popular models that work with ASTs in a different manner: TreeLSTM~\cite{DBLP:journals/corr/TaiSM15} and Code2Seq~\cite{alon2018code2seq}.
TreeLSTM is an extension of the regular LSTM~\cite{hochreiter1997lstm} network that was adapted to work with tree-structured data.
Although it falls behind in quality compared to modern models, it is useful to study since it utilizes trees without further modifications.
On the other hand, Code2Seq is a popular approach that was reused in various tasks apart from the original method name prediction~\cite{nagar2021code, Zhang2020CBPath2Vec}.
Code2Seq employs path-based representation~\cite{alon2018general} to work with code, which also relies on the information from the code's AST.

As a comparison benchmark, we use the method name prediction task for the Java programming language.
In addition to its practical applicability, method name prediction suits as a popular benchmark in the ML for SE domain~\cite{alon2018code2seq, fernandes2018structured, zugner2021codetrans}, as it allows evaluating how good models are in code understanding. We compare both models backed by eight different Java parsers. In order to do so, we develop \toolname, a tool based on PathMiner~\cite{kovalenko2019pathminer}. \toolname supports mining code representations suitable for ML models from five programming languages and twelve language-parser pairs in a unified manner.

Our results suggest that parsers can significantly affect the final quality of model predictions.
Proper parser choice for the Code2Seq model may increase the quality of the model by up to $5.5\%$ in terms of F1-score compared to the least suitable parser.
For a simpler TreeLSTM model, the parser selection is even more crucial, with the difference between the most and least suitable parsers being about $27.0\%$. Notably, for TreeLSTM and Code2Seq, the relative ordering of parser is different, meaning there are no universally good or bad parsers. Therefore, when choosing a parser, researchers and practitioners should take into account the model's properties and characteristics of trees produced by different parsers. 

With this work, we make the following contributions:
\begin{itemize}
    \item We investigate the impact of the parser selection step on the quality of machine learning models working with structural representations of code.
    To the best of our knowledge, this work is the first to study this part of the data preprocessing pipeline. We show that depending on the specific parser used, results of both TreeLSTM and Code2Seq can vary significantly.
    
    \item We present \toolname, a tool based on PathMiner, that allows running multiple parsers for different programming languages in the same manner. \toolname allows running experiments with different parsers by changing a single field in a YAML configuration file. Our tool separates the parsing step, allowing further data processing steps (\eg filtering, storage) to be parser-agnostic. It makes support of new languages, parsers, and mining tasks easy for the users.

    \item We publish a dataset of open-source Java projects called \javanorm. It is comparable in size to the commonly used \textit{Java-small} dataset, but its validation and testing parts include more diverse projects making evaluation more robust and representative.

\end{itemize}

%% file: sections/02-background.tex
\section{Background}\label{sec:background}
\newdimen\figrasterwd
\figrasterwd\textwidth

\begin{figure*}[ht]
    \centering
\parbox{\figrasterwd}{
    \parbox{.4\figrasterwd}{
        \centering
        \subcaptionbox{
Examples of rules that define syntax for method declaration.
\label{fig:rules}
        }
        {\includegraphics[width=\hsize]{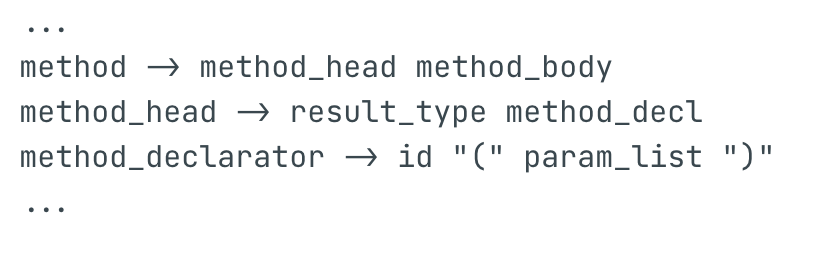}}
        \vskip3em
        \subcaptionbox{
Java method code snippet example.\label{fig:java_method}
        }
        {\includegraphics[width=\hsize]{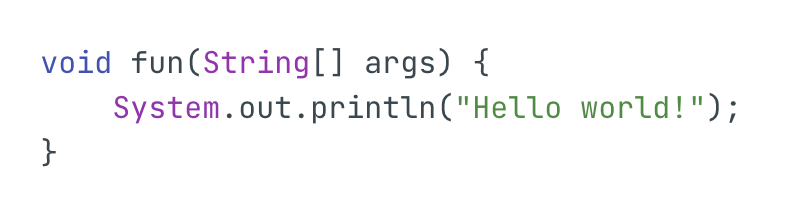}}  
    }
    \hskip1em
    \parbox{.6\figrasterwd}{%
        \subcaptionbox{
An example of a parse tree built for the code snippet on the left. Blue nodes are intermediate abstraction nodes. Green nodes contain actual code tokens.
\label{fig:ast}
        }
        {\includegraphics[width=\hsize]{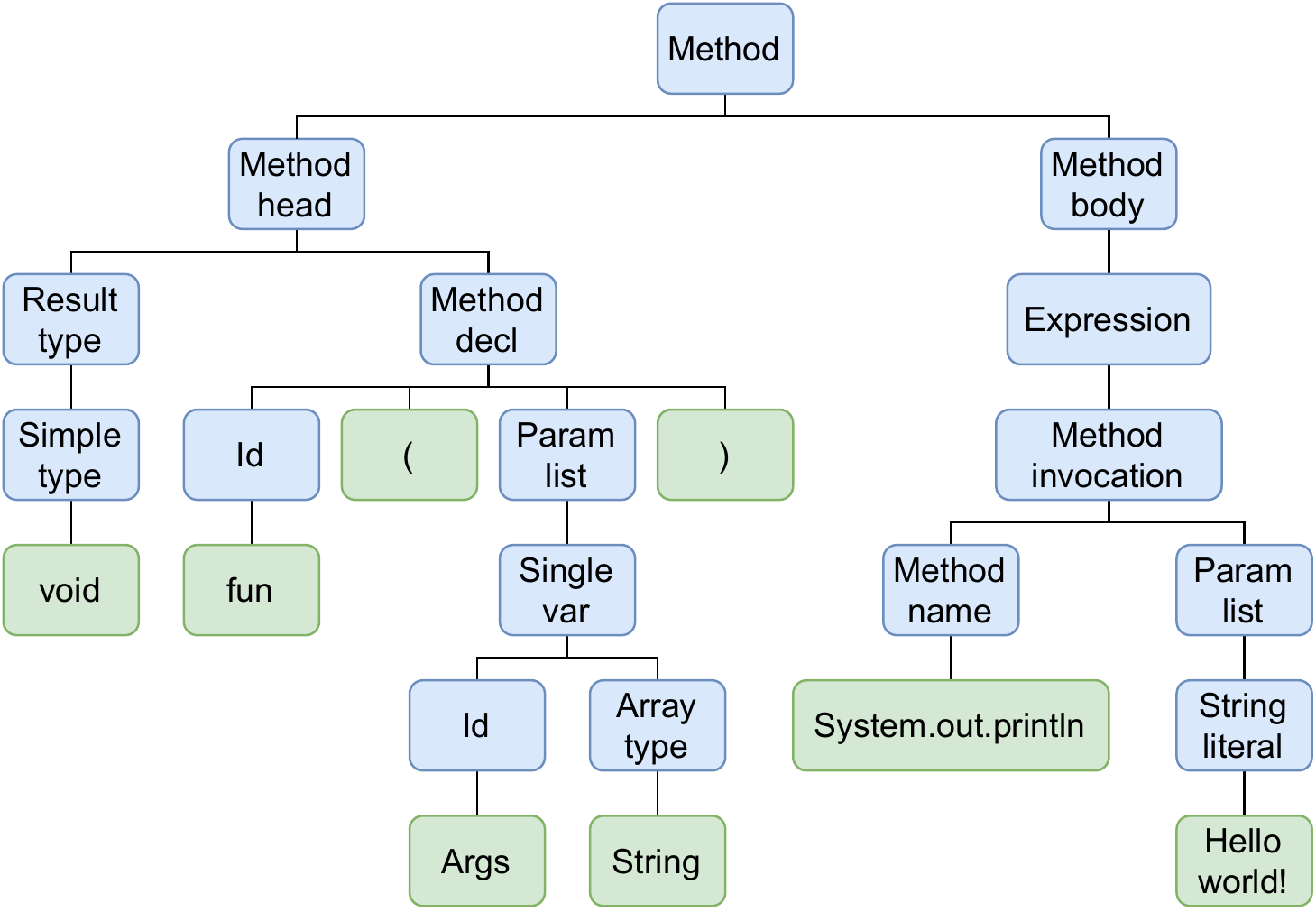}}
        }
    }
\caption{
An example of a code snippet, its correposnding parse tree, and a subset of grammar rules.
}
\vspace{-0.5cm}
\end{figure*}

In our work, we analyze the impact of choosing a parser on the quality of ML4SE models.
In this section, we first describe the main concepts of AST and then overview its applications in the ML4SE domain.

\secpart{Abstract Syntax Tree}

Each programming language has a set of rules---its grammar---that describes the language syntax.
Rules can be viewed as transformations from higher-level language abstractions into sequences of simpler abstractions and code tokens (\eg literals and variable names).
\Cref{fig:rules} shows several examples of grammar rules for method declarations in Java.

The result of parsing source code with the selected grammar is a parse tree. Each intermediate node in a parse tree represents the left-hand side of some rule, and the node's children correspond to the sequence on the right-hand side of the same rule.
Thus, intermediate nodes represent abstractions defined by the grammar and
the leaves in the parse tree correspond to tokens that actually appear in the source code.
\Cref{fig:ast} shows an example of a parse tree built for code fragment shown in \Cref{fig:java_method}.

Parse trees can be quite verbose, so most parsers can further optimize them, \eg by removing unnecessary brackets that are already reflected in the tree structure, or compressing sequences of nodes into new nodes.
After processing the parse tree, the parser constructs an abstract syntax tree (AST) that is usually more compact than the initial tree.

Parsers differ by both the grammar they use and the \linebreak post-processing behaviour.
Post-processing directly depends on the parser's application. For example, PSI~\cite{psi} and JDT~\cite{jdt} are used for language support in IDEs and, therefore, keep all information about the source code including formatting. On the opposite, JavaParser~\cite{danny_van_bruggen_2020_3842713} aims to support complex modifications of Java code, so it optimizes the tree by introducing more layers of abstraction.

\secpart{AST-based Representations in ML Models}\label{sec:ast-for-ml}

A straightforward approach to working with source code in ML models is treating code as a plain text. This approach allows applying models from the NLP domain with close to none modifications in various software engineering tasks, \eg variable misuse, code summarization.
However, as previously discussed, source code has a rich underlying structure that can be represented as an AST. Existing works in the ML4SE domain show that taking code structure into account is indeed beneficial as it leads to significant improvements in multiple SE tasks~\cite{alon2018code2seq, zugner2021codetrans, fernandes2018structured}.

Models that directly support the tree-structured input format, \eg TreeLSTM, can use raw ASTs.
While originally applied to NLP tasks, the TreeLSTM model served as a baseline in multiple ML4SE works~\cite{alon2018code2seq, chen2018tree2tree}. Shido et al~\cite{shido2019automatic} improved the model further by handling an arbitrary number of ordered children via another LSTM cell~\cite{hochreiter1997lstm}.

Another class of models builds their own code representations based on ASTs rather than using the trees as is. Alon et al.~\cite{alon2018general} suggested path-based representation of code.
Path-based representation extracts a set of path contexts from each AST, where each context is a triple of two code tokens in AST leaves and a sequence of intermediate nodes that connect these leaves.
Based on this representation, the authors suggested models for different tasks~\cite{liang2021ast, shi2020pathpair2vec}, including the Code2Vec~\cite{alon2019code2vec} and Code2Seq~\cite{alon2018code2seq} models that show high quality in method name prediction.

While AST represents the syntax of source code, a code fragment can also be represented with a graph structure. Control-flow graph (CFG) contains all possible execution paths of a program. Data-flow graph (DFG) is built from the dependencies between variables in code and their usages.
AST, DFG, and CFG can be combined into a single complex graph, as nodes from CFG and DFG also appear in the code's AST.
Allamanis et al.~\cite{allamanis2017learning} used such graph representation of code to achieve state-of-the-art results in the variable misuse task.
Lately, graph representation was used in the method name prediction~\cite{fernandes2018structured}, type inference~\cite{allamanis2020typilus}, and variable misuse~\cite{Hellendoorn2020Global} tasks to achieve superb results.

AST-based representations of code are not limited to the ones listed above.
Existing works also operate with AST traversals~\cite{Zhang2020retrieval_cs}, apply convolutions over trees~\cite{mou2016tbcnn}, enhance Transformer-based~\cite{vaswani2017transformer} models with information from ASTs~\cite{zugner2021codetrans}. 
Despite the large number of ML4SE works that employ ASTs to improve the results, they rarely motivate the choice of parsers used for AST extraction. Oftentimes, papers either omit the details about the parsing step, or just state the used tool without motivating their choice. In our work, we focus on analyzing the importance of the parser choice in ML4SE and outlining things to consider when making it.


%% file: sections/03-astminer.tex
\section{The \toolname tool}

In order to study the impact of the parser choice on the model's quality, we need a common way to extract input data for models from the source code with different parsers.
To achieve this, we need to run various parsers in a unified manner. In order to ease the maintenance and allow other researchers and practitioners to reuse our data processing pipeline, we integrate all the studied parsers in a single tool, \toolname. 

As a basis for building \toolname, we use PathMiner~\cite{kovalenko2019pathminer} --- a tool for mining ASTs and path-based representations from code in several languages, filtering the data, and storing it in a format suitable for training ML models.
The main advantage of PathMiner is that the tool is designed to be easily extendable. It provides a lot of reusable components for building parsing pipelines, which then can be used for data mining in ML4SE tasks. These components include wrappers for parsers, classes for storing AST vertices, algorithms for filtering, classes for storing the extracted trees.

By using the existing extension points, we create a pipeline that supports parsers directly imported from Kotlin and Java, parsers in other languages that run as separate processes, and grammars for ANTLR and Tree-Sitter. \toolname then runs the parsers to prepare datasets for training and evaluating ML models in the method name prediction task. Using the same pipeline, we run all parsers under equal conditions and ensure reproducibility of our experiments.

\secpart{Multilanguage Parser-agnostic Processing}

Parsers greatly vary depending on their purpose. A parser can be generated from grammar or written from scratch. It can constitute a library, or be a full-fledged runnable tool. The functionality of a parser can range from straightforward construction of parse trees to complex manipulations on source code. Finally, parsers can be written in different programming languages depending on the target applications.
Due to the variability of parsers, supporting multiple parsers in a single data preparation pipeline can be a daunting task. In \toolname, we reuse components of PathMiner and also add new extension points in order to ease the tool's usage for other researchers and practitioners.

In contrast to PathMiner, \toolname supports parsers written in languages that do not run on Java Virtual Machine (JVM). In order to add new languages and parsers, users of PathMiner can either provide a grammar for the ANTLR parser generator or implement an adapter class that wraps a third-party tool written in a JVM-compatible language. However, there are many parsers---both for Java and other languages---that do not provide Java bindings. In order to handle such cases, \toolname has a new entity called \texttt{ForeignParser}. It inherits from the \texttt{Parser} class of PathMiner and can launch standalone tools implemented in other languages to further convert their output into the inner tree structure. A similar idea was used in GumTree\cite{DBLP:conf/kbse/FalleriMBMM14}, a tool that also aggregates multiple parsers and uses them by deserializing their XML output. In our implementation, we use JSON, as this format is more common nowadays.

PathMiner could be easily extended to support new languages by using automatically generated parsers based on ANTLR grammar files. Following this idea, we also add the support of Tree-sitter~\cite{max_brunsfeld_2022_6326492} grammars. Due to the ease of usage, the availability of Python bindings, and the support of many programming languages, Tree-Sitter is commonly being used in ML4SE data mining pipelines. 

Thus, \toolname currently supports:
\begin{enumerate}
    \item parsers that provide JVM bindings;
    \item parsers that can be run as stand-alone processes or wrapped in executable programs with console output;
    \item parsers derived from ANTLR and Tree-sitter grammars.
\end{enumerate}
Importantly, each of these methods has the same API and the work of the parsers is completely hidden from the rest of the \toolname components. For now, our tool already supports 5 different languages with 12 parser/language pairs. While the parser support can be extended, in this study we mainly focus on Java language as we selected it for the evaluation of models.



\secpart{Data-processing Pipeline}

\begin{figure*}
    \centering
    \includegraphics[width=\textwidth]{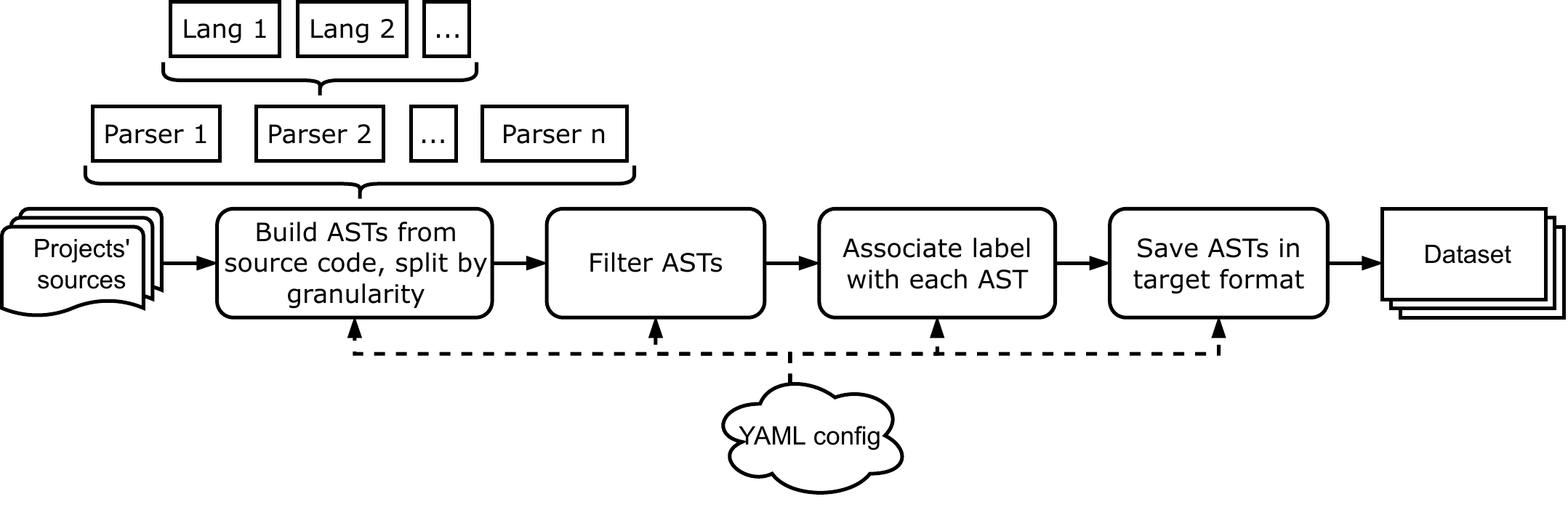}
    \caption{
Overview of the data processing pipeline in \toolname. Users can configure the pipeline to suit their needs by providing a YAML configuration file. For example, user can configure the tool to do the following: (1) parse Java code with ANTLR parser, (2) extract ASTs of all methods, (3) remove overriden methods, (4) associate each tree with method name, and (5) store the resulting dataset in JSON Lines format.
}
    \label{fig:astminer-pipeline}
\end{figure*}

In \toolname, we reuse and improve many components from PathMiner. We combine them into a single pipeline that can be launched from the command line and configured by providing a YAML file.
Novel approaches often rely on large-scale data analysis, thus, we significantly speed up the tool by adding multithreading processing and provide an excessive logging system together with the command-line interface, making it convenient to use \toolname on remote servers.
 

After the parsing stage, \toolname generates a single tree or a set of trees from each file depending on the selected granularity (file or method). The trees have a unified parser-independent form and can then be processed depending on the target task. In order to make \toolname easily extensible to new tasks and ML models, the following processing pipeline consists of several steps which can be configured and changed independently. 
All these steps are represented by simple programming interfaces that could be implemented to extend \toolname with new behaviour. Along with the source code of the tool, we also publish an extensive documentation describing its main components.

\textbf{Label extraction}. Label extractors associate each tree with some string (a label) which can then be used as a target for model training and evaluation. This component of the pipeline can be replaced depending on the task. \toolname currently supports extraction of method names, file names, and parent folder names.

\textbf{Filtering}. At the next stage, \toolname applies filters to all the extracted trees and labels. Label filters can be used to remove samples with the desired labels that cannot be supported by a model, \eg names in unsupported languages, too long or too short names. Tree-level filters can be either structural or content-based. Structural filters remove samples based on the AST characteristics: depth, number of vertices. Content-based filters operate with the tree itself and filter samples by annotations (\eg removing overridden methods or tests), modifiers (\eg removing abstract methods), and the size of the method's body.

In order to avoid implementing a separate set of filters for each parser, we accompany parsed trees with a set of features pre-computed at the parsing stage. This set includes both external and internal features. External features include names of the file and enclosing elements (\eg classes and enumerations). Internal features are function parameters, its name, modifiers, annotations, and others. These features are stored in a parser-independent form, which allows users to implement new filters without manual analysis of the parsed structure. 

\textbf{Storing}. At the last stage of the pipeline, we convert the trees to the selected format and finally save the dataset on disk. For storing the data, we mostly reuse components of PathMiner, but also add the ability to compress trees before storing. In addition, we save metadata alongside with the mined dataset, such as the origin of the tree (\ie path to the enclosing file) or the location of the corresponding code snippet in the file.

To compress trees, we use the algorithm presented by Yin et al.~\cite{yin2017syntactic}. The authors use production rule closures and concatenate the names of the rules into a single vertex type. In order to make the process parser-independent, we imitate this algorithm by compressing ``bamboos'', \ie sequences of nodes in a tree that have a single child each, into single nodes in the already generated AST.


%% file: sections/04-experiments.tex
\section{Experimental setup}\label{sec:motivation}

\secpart{Research Questions}
With this study, we aim to investigate the importance of the initial source code parsing step for the machine learning on source code algorithms.
Moreover, we focus on the differences of how source code is parsed and how these differences affect the final performance of ML models.
We formulate the following research questions:

\begin{itemize}
    \item \textbf{RQ1:} Is there any statistical difference in the ASTs produced by different parsers?
    \item \textbf{RQ2:} How does the parser selection affect machine learning models that take into account tree structure of code?
    \item \textbf{RQ3:} Alongside with the model quality, what else researchers and practitioners should take into account when choosing a parser for their data-processing pipeline?
\end{itemize}

With RQ1, our goal is to understand whether different parsers indeed lead to varying characteristics of model input.


By answering RQ2, we aim to understand whether the differences in trees produced by different parsers lead to changes in the quality of machine learning models.
If the answer is positive, further research should pay attention to the choice of a parser in order to maximize the performance of the studied models and to ensure fair comparison.

In RQ3, we address practical features of code parsers, such as their speed, size of the generated datasets, and language support. These aspects become important when choosing between parsers that allow models to achieve similar quality.



\secpart{Parsers}
As a target programming language for this study, we use Java due to its common adoption in the ML4SE domain. Java is a fairly common target for syntactic and semantic analysis, as well as for research works on code summarization~\cite{alon2018code2seq, fernandes2018structured}, code generation~\cite{wei2019cg-cs}, and others~\cite{liang2021ast}.
Also, due to the language's popularity, long history of development, and numerous applications, there exist a lot of different parsers for Java.

There are many tools that convert source code into a tree structure. The main examples are interpreters and compilers that use it to make the code executable. Other examples are refactoring and code highlighting systems (\eg PSI trees in the IntelliJ Platform), algorithms for comparing source code~\cite{DBLP:conf/kbse/FalleriMBMM14}, algorithms for finding duplicates in code~\cite{tairas2006phoenix}, and other tasks where program structure is important. Due to the variety of applications, there are many parsers for different scenarios.
For this work, we selected the following parsing tools and libraries:
    
\textbf{ANTLR}~\cite{10.5555/2501720} --- a parser generator that takes a grammar for the target language as input and generates a parser implementation in one of the supported languages. In this work, we use ANTLR with an open-source grammar for Java.\footnote{Java 8 ANTLR grammar: \url{https://github.com/antlr/grammars-v4/tree/master/java/java8}}

\textbf{Eclipse Java Development tools (JDT)}~\cite{jdt} --- a set of plugins for Eclipse IDE that are responsible for supporting Java. Based on JDT, the IDE can highlight Java code, refactor it, perform different checks, and more.
We use GumTree~\cite{DBLP:conf/kbse/FalleriMBMM14} Java API to add JDT support to \toolname.   

\textbf{JavaLang}~\cite{tunes} --- an open-source library written in pure Python for working with the Java 8 source code. Due to the implementation in Python, JavaLang can be easily integrated into ML and data science pipelines. The library focuses on parsing and does not provide tools for further manipulations with the extracted trees.

\textbf{JavaParser}~\cite{danny_van_bruggen_2020_3842713} --- a Java library for parsing, transformation, and generation of Java code. It allows analyzing Java programs, generating boiler-plate code, and focuses on complex manipulations with the generated ASTs. Notably, it has been used in some of the previous ML4SE studies~\cite{alon2018code2seq, zugner2021codetrans,fernandes2018structured}.

\textbf{Program Structure Interface (PSI)}~\cite{psi} --- similar to JDT, it is a tool that is responsible for the language support in IDEs built upon the IntelliJ Platform. It allows manipulation of source code inside the IDEs and their plugins, running inspections, refactorings, and more.
We use PSIMiner~\cite{psiminer} to extract PSI trees from code.

\textbf{Spoon}~\cite{pawlak:hal-01169705} --- an open-source library for parsing, rewriting, transforming, and transpiling Java code. In contrast to previous tools, it was developed in the research community and since then has been used in numerous works on refactoring and program repair~\cite{xuan2016nopol, martinez2016astor, long2017automatic}.

\textbf{SrcML}~\cite{collard2013srcml} --- a lightweight tool for analyzing and manipulating source code. It allows transforming C, C++, and Java code by using trees in a native SrcML format inspired by XML. The authors provide a set of tools for further manipulation of the extracted trees: static type resolution, pointer analysis, builder of UML diagrams, etc.
To support this parser in \toolname, we also use GumTree.

\textbf{Tree-Sitter}~\cite{max_brunsfeld_2022_6326492} --- a parser generator focused on the usage in text editors and IDEs in order to provide features like syntax highlighting in real time. Due to the need for real-time usage, Tree-sitter can parse code incrementally, is fast and robust when parsing unfinished code. Similar to ANTLR, we use its most popular grammar for Java 8.\footnote{Java 8 Tree-sitter grammar: \url{https://github.com/tree-sitter/tree-sitter-java}}

\smallskip

The chosen parsers target different applications. Each parser has its own features: from being lightweight and fast, to supporting complex manipulations of ASTs. This diversity allows us to assess whether ASTs extracted with different parsers are similar or they can affect the quality of ML models working with source code.

\secpart{Models}

In this work, we compare two different ML models, TreeLSTM~\cite{DBLP:journals/corr/TaiSM15} and Code2Seq~\cite{alon2018code2seq}, that can solve software engineering problems by working with tree-based representation of code or its extensions.
With this, we aim to make a diverse selection from the point of model architecture and tree input format.

\textbf{TreeLSTM.} Originally created for natural language processing, the TreeLSTM model was then reused and adapted for working with source code~\cite{alon2018code2seq}.
Although this model is simpler and shows inferior quality compared to the novel approaches, TreeLSTM works directly with tree representation of code.
This is crucial for our research, since it allows comparing the impact of parsing differences directly, without additional transformations of trees (\ie adding new edges in DFG or CFG~\cite{allamanis2017learning}).



\textbf{Code2Seq.} On the other side, Code2Seq is a popular model showing good results in many SE applications~\cite{alon2018code2seq, nagar2021code, Zhang2020CBPath2Vec}.
This model receives information about source code from a path-based representation, \ie a set of contexts where each context is a triple of tokens in two leaves in a tree and a sequence of intermediate nodes on a path between them. As Code2Seq does not work directly with ASTs, we can assess whether the choice of a parser can affect the models that use code representations derived from the ASTs.



\secpart{Dataset}\label{sec:dataset}

In order to ensure correctness of the results and reduce randomness, we run each model five times with different random seeds for each parser. Thus, train and evaluate models $80$ times in total (2 models $\times$ 5 random seeds $\times$ 8 parsers). Due to the high number of experiments, initially we decided to use the \emph{Java-small} dataset~\cite{allamanis2015acc_method_names}. In our case, using larger datasets such as \emph{Java-med} and \emph{Java-large} is too computationally intensive as they are 5 to 30 times larger than \emph{Java-small}.

However, our experiments and previous experience with \emph{Java-small} suggest that evaluating the models on this dataset comes with two important limitations:
\begin{enumerate}
    \item The models' quality reported on the validation data weakly correlates with the results on the testing data and is often way lower.
    It makes tuning of hyper-parameters based on the validation quality unreliable.
    \item Model evaluation on \emph{Java-small} is not robust, with small model modifications leading to large deviations in the final metrics. As a result, evaluation on this dataset often disagrees with larger datasets in terms of relative model ranking.
\end{enumerate}

Our investigation of these issues suggests that a significant disadvantage of \emph{Java-small} is that both its validation and testing parts consist of a single project each.
Judging models quality by evaluating on a single project might lead to non-representative results since every project is unique in some way.
Moreover, the \texttt{libgdx} project that constitutes the validation set is not a typical Java project, with more than 45\% of code written in C and C++, which might explain the discrepancy between the validation and testing results. 

In order to deal with this issue, we compile a new dataset called \javanorm that has a size comparable to \emph{Java-small}, but aims to mitigate its issues. \javanorm contains five projects in both validation and testing parts (see \Cref{table:data_intersected} for the list of projects). The training part of the new dataset includes ten open-source Java projects. As the set of projects in both testing and validation is more diverse, evaluation on \javanorm is more representative and robust.

We collect projects that are publicly available on GitHub.\footnote{GitHub: \url{https://github.com/}} 
While collecting the data, we respect all intellectual rights and select only the projects whose licenses allow to reuse the code for further analysis.
We collect projects that have one of the following licenses: Apache-2.0, MIT, or BSD-3.0.
We consider projects that are not forks and have at least 20MB of Java code.
Across all projects that meet these criteria, we take the ones with the highest numbers of GitHub stars.

We make \javanorm publicly available via Zenodo.\footnote{\javanorm: \url{https://doi.org/10.5281/zenodo.6366599}} As it mitigates some issues of \emph{Java-small} while being comparable in size, we believe that the dataset can be useful for future research in the method name prediction and other ML4SE tasks.

\secpart{Evaluation Scheme}

In the rest of this section, we describe the evaluation scheme we use to evaluate the impact of parsers on models' quality and answer the research questions.

\textbf{Tree comparison.}
In order to verify whether trees built by different parsers are similar or not, we compare them by several metrics: tree depth, size, branching factor, etc. The values of a certain metric for different trees produced by the same parser can be viewed as a distribution of an unknown form.
Therefore, to compare trees produced by different parsers, we can compare the distributions of their metrics.

There exist multiple approaches to compare numerical distributions.
One of the most popular is the Student's t-test~\cite{student1908probable} that tests the hypothesis of mean equality between Normal distributions.
Although values of a certain metric for a single parser are not normally distributed, with a dataset of 300 thousand trees, we can apply the Central Limit Theorem and work with this test.

\textbf{Method name prediction.}
As a benchmark for our study, we select the method name prediction problem.
More formally, the model should generate a name of the method given its body and signature with the original name being replaced with a stub token.
This task is a common benchmark for analyzing model capabilities in code understanding and summarization~\cite{alon2018code2seq, fernandes2018structured, zugner2021codetrans}.


As a de-facto standard, we treat the method name prediction task as a sequence generation problem.
In this setting, each method name is split into a sequence of sub-tokens by CamelCase or snake\_case.
Models then generate sub-tokens one by one, taking the code snippet and already predicted sub-tokens into account.

\textbf{Intersection of datasets.}\label{sec:dataset-intr}
As different parsers use different underlying algorithms, they might parse different sets of files or methods from the same code corpus. It might happen due to several reasons: some parsers fail for specific files due to the presence of unexpected symbols, syntactically incorrect or incomplete code, unsupported versions of Java, etc. Also, processing of trees for each parser requires significant manual parser-specific work (\eg to correctly split the trees into methods or extract modifiers) which might fail in corner cases. As a result, the set of parsed ASTs from the same initial code dataset slightly differs from parser to parser.

In order to correctly answer our research questions, the resulting datasets for all the parsers must contain trees for exactly the same set of methods. Otherwise, the diversity in the models' quality may be attributed to the differences in the training and testing sets.
In order to mitigate this, we \textit{intersect datasets}, \ie  we keep only those methods that may be successfully parsed by each parser. For this, we store a source range for each extracted method --- the exact position where this method appears in the source code. Based on the source ranges, we match methods extracted by different parsers. In total, we dropped less than 1\% of the methods after the intersection, \Cref{table:data_intersected} provides information about the final dataset.

\begin{table}[htbp]
    \centering
 \begin{tabular}{c|ccc}
 \toprule
                           & \textbf{Training}   & \textbf{Validation}     & \textbf{Testing}             \\ \hline

\multirow{10}{*}{\textbf{Projects}} & cassandra        & \multirow{2}{*}{buck}   & \multirow{2}{*}{bazel}    \\
                           & gocd             &                         &                           \\
                           & gradle           & \multirow{2}{*}{dl4j}   & \multirow{2}{*}{dbeaver}  \\
                           & hbase            &                         &                           \\
                           & kafka            & \multirow{2}{*}{druid}  & \multirow{2}{*}{guava}    \\
                           & pentaho-kettle   &                         &                           \\
                           & presto           & \multirow{2}{*}{flink}  & \multirow{2}{*}{keycloak} \\
                           & pulsar           &                         &                           \\
                           & spring-framework & \multirow{2}{*}{j2objc} & \multirow{2}{*}{quarkus}  \\
                           & tomcat           &                         & \\ \hline
\begin{tabular}[c]{@{}c@{}}\textbf{Number of}\\ \textbf{methods}\end{tabular}                & 295,425          & 187,717 & 144,656                  \\ 
\bottomrule
\end{tabular}
    \caption{Sets of projects and number of methods (after the intersection step) in training, validation, and testing parts of \javanorm.}
    \label{table:data_intersected}
\end{table}

\textbf{Metrics.}
The most popular approach to measure quality of method name prediction is F1-score~\cite{allamanis2015acc_method_names}.
Originally applied for classification, it was adopted to evaluate the quality of generated sequences. F1-score is a harmonic mean of precision and recall.
Given two sequences, ground truth and prediction, precision is a proportion of sub-tokens from the predicted sequence that appear in the ground truth.
Recall, as an opposite, is a proportion of correct sub-tokens that appeared in the predicted sequence.
Finally, F1-score is a harmonic mean of these two values.

Although F1-score is a popular metric, it also has disadvantages, \eg it does not take into account the order of sub-tokens. While for some method names (\eg \texttt{arraySort} and \texttt{sortArray}) it does not affect the meaning, in other cases it might alter it (\eg \linebreak \texttt{implementationParser} and \texttt{parserImplementation}).
Recently, Roy et al.~\cite{roy2021reassessing} evaluated the applicability of multiple metrics from the NLP domain in code summarization tasks. Specifically, the authors evaluated METEOR~\cite{banerjee2005meteor}, BLEU~\cite{papineni2002bleu}, Rouge~\cite{lin2004looking}, ChrF~\cite{popovic2015chrf}, and BERTScore~\cite{zhang2019bertscore}.
Judging by their results, ChrF~\cite{popovic2015chrf} appears to be the most robust metric for code summarization, so we employ it in our study alongside the F1-score.

\textbf{Model comparison.}
The answer to RQ2 requires pairwise comparisons of independent models trained and tested on the same set of methods, but with data prepared by different parsers.
To do so, we run the paired bootstrap test~\cite{Efron93bootstrap} over the predictions of two models.
The test resamples the result pairs with replacement multiple times and compares average results of the models on the sampled examples.
The probability of one model beating the other is a ratio of how often the average score for this model is better than for the other to the total number of runs.

%% file: sections/05-results.tex
\section{Results}\label{sec:results}

We use the described evaluation scheme for per-parser analysis of the produced ASTs and further comparison of two ML4SE models: TreeLSTM~\cite{DBLP:journals/corr/TaiSM15} and Code2Seq~\cite{alon2018code2seq} on the method name prediction problem. We use a newly collected \javanorm dataset, leaving only the methods successfully processed by all the studied parsers.

\textbf{RQ1: Is there any statistical difference in the ASTs produced by different parsers?}\label{sec:tree_analysis}

\begin{table*}[]
    \centering
    \input{tables/parser-diff}
    \caption{
On the left-hand side: median values of the computed metrics for each parser. TS -- tree size, TD -- tree depth, BF -- branching factor, UTP -- unique types, UTK -- unique tokens.
On the right-hand side: parsers' features that can be considered when building a data processing pipeline.
\\
$^*$We extract PSI trees with PSIMiner~\cite{psiminer} rather than with \toolname. PSIMiner is slower compared to \toolname as it opens each project in IntelliJ IDEA, which introduces a significant time overhead.
    }
    \label{table:tree-stats}
\end{table*}

\begin{table*}[]
    \centering
    \input{tables/ttest}
    \caption{Tree metrics that are similar for trees produced by each pair of parsers according to the Student's t-test with $p$-value 0.01. For example, trees built by ANTLR and Spoon are similar w.r.t. tree depth. 
    }
    \label{table:ttest}
\end{table*}

In order to answer the first research question, for each tree, we compute the following set of metrics:
\begin{itemize}
    \item \textbf{tree size} (TS) --- number of nodes in a tree;
    \item \textbf{tree depth} (TD) --- number of nodes in a path from the root of a tree to its deepest node;
    \item \textbf{branching factor} (BF) --- mean number of children in non-leaf vertices of a tree;
    \item \textbf{unique types} (UTP) --- number of unique types of intermediate nodes used in an AST, lower values correspond to a higher level of abstraction used in a parser as it can represent the same code fragment in a more compact way;
    \item \textbf{unique tokens} (UTK) --- number of unique sub-tokens in AST leaves, lower values also reflect higher level of abstraction (\eg whether parser keeps binary operators as tokens or as node types).
\end{itemize}

The first three metrics aim to reflect the tree structure and estimate how different the trees are size-wise.
The latter two metrics analyze the content and abstraction level of trees produced by different parsers.
\Cref{table:tree-stats} provides quantitative information about the computed metrics for each parser.

After computing the selected five metrics for trees built by each parser, we apply the Student's t-test~\cite{student1908probable} to determine whether the distributions differ for each pair of parsers.
We consider distributions to be significantly different if the corresponding \textsc{p-value} is less than $0.01$, \ie we consider distributions different only if the probability of a mistake is less than 1\%.
According to the conducted statistical tests, all parser pairs produce trees different by at least three metrics. \Cref{table:ttest} shows with respect to which metrics the output of two parsers is similar for each pair of parsers.

\begin{tcolorbox}[enhanced jigsaw,colback=bg,boxrule=0pt,arc=0pt]
\textbf{Take away 1.} By applying different parsers to the same code fragment we get different trees both in terms of structure (\ie tree size, depth, and branching) and in terms of content (\ie the number of unique types and tokens in the trees). Although some parsers produce trees that are similar by one or two metrics, for all parser pairs trees differ significantly in various ways.
\end{tcolorbox}

\textbf{RQ2: How does the parser selection affect machine learning models that take into account tree structure of code?}

Since RQ1 shows that different parsers produce trees with different characteristics, it is necessary to study whether it affects the quality of models that utilize these trees. 
For that, we train both models, TreeLSTM and Code2Seq, on the datasets produced by each parser.
To get reliable results, we train each model-parser pair five times with different random seeds.
Therefore, we train and evaluate $80$ models overall. \Cref{table:model-results} shows mean values and variance of metrics for each model-parser pair.

\begin{table*}[]
    \centering
    \input{tables/model-results}
    \caption{Quality of method name prediction for TreeLSTM and Code2Seq models on the data extracted with the studied parsers.}
    \label{table:model-results}
    \vspace{-0.2cm}
\end{table*}

\newdimen\figrasterwd
\figrasterwd\textwidth
\begin{figure*}[ht]
    \centering
    \parbox{\figrasterwd}{
        \centering
        \subcaptionbox{
Paired bootstrap for TreeLSTM model.\label{fig:bootstrap-treelstm}
}
        {\includegraphics[width=0.77\textwidth]{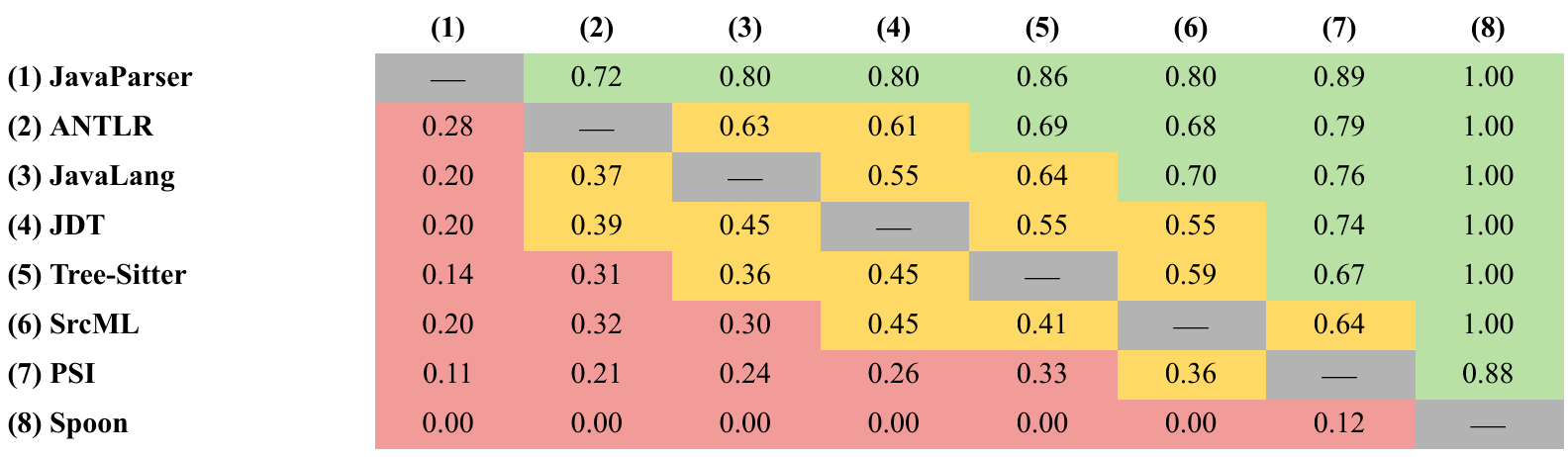}}
        \subcaptionbox{
Paired bootstrap for Code2Seq model.\label{fig:bootstrap-code2seq}
        }
        {\includegraphics[width=0.77\textwidth]{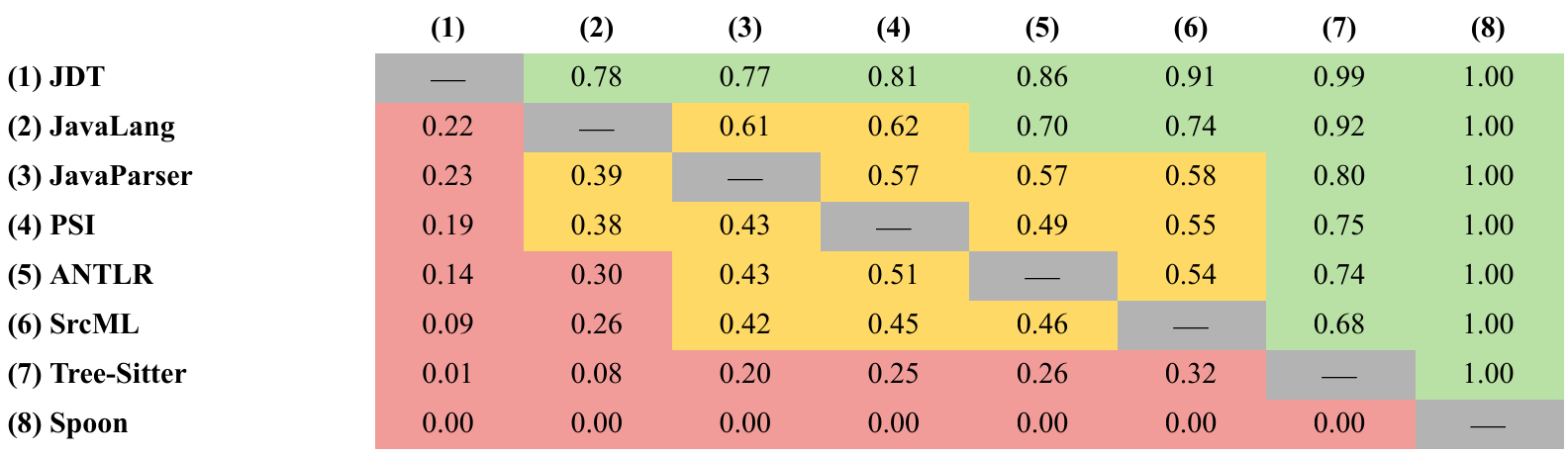}}
    }
\caption{
Results of pairwise bootstrap tests for TreeLSTM and Code2Seq models. The numbers are probabilities of one model beating another with random initialization and on a random subset of data. Colors are added for clarity.\label{fig:bootstrap}
}
\end{figure*}

The direct comparison already shows the difference between parsers for both models.
An accurate parser choice can increase F1-score by up to $27\%$ for TreeLSTM and $5.5\%$ for Code2Seq compared to the worst performing parser. For ChrF the results are similar, with $31\%$ and $8\%$ increase for TreeLSTM and Code2Seq, respectively.
While the absolute differences might seem low, we argue that they cannot be ignored in current ML4SE research. Nowadays, most works compare the models by directly comparing metric values averaged over the testing dataset (average F1-score in the case of method name prediction). For example, in the recent works on method name prediction~\cite{zugner2021codetrans,fernandes2018structured}, the differences of less than 2 percentage points in F1-score between different models and different model variations are considered significant, while some of the compared models heavily rely on code structure. Thus, the observed differences in metrics depending on the parser are comparable to the values considered significant in modern works.

In order to further ensure correctness of the results, we also conduct paired bootstrap tests for the trained models with respect to F1-score.
For each pair of parsers and random seeds, we run their pairwise comparison with 1,000 bootstrap resamples, 25,000 resamples in total (5 random seeds for both parsers). We then estimate the probability of one parser beating another as the percentage of resamples where it has a higher average score.
\Cref{fig:bootstrap-treelstm} and \Cref{fig:bootstrap-code2seq} provide the computed probabilities for both models and all analyzed parsers. Given the current approach to model comparison (\ie direct comparison of averaged metrics over the testing dataset), the values can be interpreted as the probability of the same model with one parser showing better quality than with another one.


Firstly, the bootstrap results suggest that parsers indeed have an impact on the model quality. Otherwise, all the pairwise winning probabilities would be close to 50\%. Secondly, for both models, there exists a parser (JavaParser for TreeLSTM, JDT for Code2Seq) that beats the others most of the time and can be treated as a superior choice when working with this model. Also, there are parsers for each model (\eg ANTLR, JavaLang, and JDT in the case of TreeLSTM) that show similar performance to each other and can be used interchangeably depending on the existing limitations (\eg speed, ease of integration into the mining pipeline). Finally, for the two studied models, the relative ordering of parsers differs. While for TreeLSTM JavaParser is the best choice, and Spoon and PSI show poor quality, for Code2Seq it is JDT which is the best, and Spoon and Tree-sitter are falling behind the others. 

Combining the changes in models' quality with the tree metrics computed in RQ1 allows us to make assumptions about the nature of the changes.
We notice that shallower trees generally lead to better model performance, which is intuitive since ML models may struggle to pass information on long distances.
Another observaion is that the small size of the trees combined with a higher level of their abstraction (fewer unique types and tokens) helps JavaParser, JavaLang, and JDT consistently outperform the others.

\begin{tcolorbox}[enhanced jigsaw,colback=bg,boxrule=0pt,arc=0pt]
\textbf{Take away 2.} The choice of a parser for data preprocessing impacts the model's quality.
Although some parsers show dominance over others, the optimal choice is specific to each model.
Without additional information about the model, parsers that produce compact trees with higher level of abstraction are more preferable. 
\end{tcolorbox}

\textbf{RQ3: Alongside with the model quality, what else \linebreak researchers and practitioners should take into account when choosing a parser for their data-processing pipelines?}



For both models, there exist groups of parsers showing similar performance. When choosing between such parsers, aspects other than raw model quality should be taken into consideration.
Depending on the task at hand, the size of the dataset, and target programming languages, different parsers might become more or less suitable. 

When dealing with large-scale datasets, the parser's speed might become a bottleneck. \Cref{table:tree-stats} shows the average number of processed examples per second for all the parsers.
To measure this, we run \toolname multiple times for each parser with a task of preparing data for the TreeLSTM model, for which parsing is the most time-consuming stage of the pipeline.
The fastest parsers considered in this study---JDT and JavaParser---are more than 8 times faster than JavaLang and more than 6 times faster than Tree-sitter.

Another resource-related issue is the memory required to store the resulting dataset. It might affect the speed of model training: smaller datasets are faster to read from drive, and in some cases can be loaded directly into RAM. Memory consumption directly depends on the size of the produced trees. The Spoon and PSI parsers generate 2 to 4 times larger trees compared to others. The TS column in \Cref{table:tree-stats} provides average tree sizes for different parsers, which allows to approximate the size of the processed dataset.

Support of multiple programming languages allows the whole learning pipeline to be less dependent on a specific language and thus be more extensible. Moreover, it is necessary for training multi-language models~\cite{feng2020codebert}. Parser generators like ANTLR\footnote{ANTLR supported languages: \url{https://github.com/antlr/grammars-v4}} and Tree-Sitter\footnote{Tree-Sitter supported languages: \url{https://tree-sitter.github.io/tree-sitter/\#available-parsers}} can support any language if the appropriate grammar is available, making them the most flexible choice. In addition to Java, SrcML also supports C, C++, and C\#.
While, PSI works for all the languages supported by IntelliJ-based IDEs and their plug-ins, extraction of PSI for each language requires manual work to extend PSIMiner. Currently, PSIMiner supports Java and Kotlin.
The rest of the tools that we consider support only the Java language.

Finally, since most ML pipelines use Python as the main language, Python bindings might be convenient to interact with the parser directly. Out of the considered parsers, only ANTLR, Tree-Sitter, and JavaLang provide the necessary bindings, while the other can only be used with JVM.

\begin{tcolorbox}[enhanced jigsaw,colback=bg,boxrule=0pt,arc=0pt]
\textbf{Take away 3.} In addition to the parser's impact on the model quality, there are other aspects that researchers and practitioners might consider while selecting a parser: speed, language support, ease of usage.
The exact limitations on the parser choice come from the studied problem, usage scenarios, and the selected model architecture. In order to choose a parser for data processing, researchers should first identify parsers that are applicable in their situation, and then treat the applicable parsers as a hyperparameter. The final choice can be made via empirical validation or by carefully studying features of trees produced by each parser and the nuances of a particular ML model. 
\end{tcolorbox}



%% file: tables/parser-diff.tex
\begin{tabular}{l|ccccc|ccc}
\toprule
\multirow{2}{*}{\textbf{Parser}} &
    \multicolumn{5}{c|}{\textbf{Tree metrics (median)}} &
    \textbf{Language} & \textbf{Python} & \textbf{\toolname} \\
& \textbf{TS} & \textbf{TD} & \textbf{BF} & \textbf{UTP} & \textbf{UTK} &
\textbf{support} & \textbf{API} & \textbf{speed (files/sec)}
\\
\hline
\textbf{ANTLR} & 56 & 8 & 3 & 32 & 24 & Any (grammar) & + & 759 \\
\textbf{JDT} & 31 & 6 & 2 & 14 & 19 & Java & - & 1045 \\
\textbf{JavaLang} & 36 & 7 & 2 & 19 & 19 & Java & + & 78 \\
\textbf{JavaParser} & 28 & 6 & 2 & 13 & 17 & Java & - & 938 \\
\textbf{PSI} & 93 & 9 & 2 & 32 & 29 & Any (IntelliJ) & - & 30$^*$ \\
\textbf{Spoon} & 109 & 8 & 2 & 15 & 28 & Java & - & 150 \\
\textbf{SrcML} & 42 & 8 & 2 & 17 & 19 & Java/C/C++/C\# & - & 313 \\
\textbf{Tree-Sitter} & 60 & 8 & 3 & 25 & 27 & Any (grammar) & + & 117 \\
\bottomrule
\end{tabular}

%% file: tables/ttest.tex
\begin{tabular}{l|cccccccc}
\toprule
{} & \textbf{ANTLR} & \textbf{JDT} & \textbf{JavaLang} & \textbf{JavaParser} & \textbf{PSI} & \textbf{Spoon} & \textbf{SrcML} & \textbf{Tree-Sitter} \\
\hline
\textbf{ANTLR}       & - & - & - & - & UTP & TD & TD & TS \\
\textbf{JDT}         & - & - & TS, UTK & TS, UTK & - & - & UTK & - \\
\textbf{JavaLang}    & - & TS, UTK & - & UTK & - & - & TS, UTK & TD  \\
\textbf{JavaParser}  & - & TS, UTK & UTK & - & BF & - & UTK & - \\
\textbf{PSI}         & UTP & - & - & BF & - & UTK & - & UTK \\
\textbf{Spoon}       & TD & - & - & - & UTK & - & TD & UTK  \\
\textbf{SrcML}       & TD & UTK & TS, UTK & UTK & - & TD & - & - \\
\textbf{Tree-Sitter} & TS & - & TD & - & UTK & UTK & - & - \\
\bottomrule
\end{tabular}

%% file: tables/model-results.tex
\sisetup{detect-weight,mode=text}
\renewrobustcmd{\bfseries}{\fontseries{b}\selectfont}
\renewrobustcmd{\boldmath}{}
\newrobustcmd{\B}{\bfseries}

\begin{tabular}{l|cccc|cccc}
\toprule
\multirow{2}{*}{\textbf{Parser}} &
    \multicolumn{4}{c|}{\textbf{TreeLSTM}} &
    \multicolumn{4}{c}{\textbf{Code2Seq}} \\

{} & \textbf{Prec} & \textbf{Rec} & \textbf{F1} & \textbf{ChrF} &
     \textbf{Prec} & \textbf{Rec} & \textbf{F1} & \textbf{ChrF} \\

\hline
\textbf{ANTLR} & $36.0 \pm 1.9$ & $25.5 \pm 1.1$  & $29.8 \pm 1.3$ & $24.8 \pm 1.4$ & $50.2 \pm 1.1$ & $32.7 \pm 0.7$ & $39.6 \pm 0.4$ & $31.3 \pm 0.6$ \\
\textbf{JDT}   & $35.9 \pm 2.2$ & $24.5 \pm 1.3$ & $29.1 \pm 1.6$ & $25.0 \pm 1.2$ & \textbf{51.0 $\pm$ 0.6} & \textbf{33.4 $\pm$ 0.5} & \textbf{40.3 $\pm$ 0.3} & $32.0 \pm 0.4$ \\
\textbf{JavaLang} & $36.4 \pm 1.9$ & $24.7 \pm 1.0$  & $29.4 \pm 1.0$ & $25.1 \pm 1.3$  & $50.7 \pm 0.8$ & $33.1 \pm 0.5$  & $40.0 \pm 0.1$    & $31.9 \pm 0.4$ \\
\textbf{JavaParser} & \textbf{38.2 $\pm$ 2.0} & \textbf{26.0 $\pm$ 1.4} & \textbf{30.9 $\pm$ 1.6} & \textbf{25.7 $\pm$ 1.8} & $49.7 \pm 0.9$ & $33.3 \pm 0.3$ & $39.9 \pm 0.3$  & \textbf{32.2 $\pm$ 0.1} \\
\textbf{PSI} & $35.6 \pm 3.7$ & $22.9 \pm 1.7$ & $27.8 \pm 2.3$ & $23.3 \pm 1.3$  & $50.3 \pm 0.6$  & $32.9 \pm 0.5$  & $39.8 \pm 0.4$ & $31.7 \pm 0.3$ \\
\textbf{Spoon} & $31.6 \pm 1.9$ & $19.9 \pm 1.7$  & $24.3 \pm 1.8$ & $19.6 \pm 1.2$ & $50.1 \pm 0.8$ & $30.9 \pm 0.5$ & $38.2 \pm 0.5$ & $29.8 \pm 0.3$ \\
\textbf{SrcML} & $36.1 \pm 1.9$  & $24.1 \pm 2.0$ & $28.9 \pm 2.0$ & $24.3 \pm 1.8$ & $50.3 \pm 0.9$ & $32.7 \pm 0.4$ & $39.7 \pm 0.5$ & $31.5 \pm 0.2$ \\
\textbf{Tree-Sitter} & $36.4 \pm 1.4$ & $24.1 \pm 1.0$ & $29.0 \pm 1.1$ & $24.0 \pm 0.3$ & $50.4 \pm 0.5$ & $32.4 \pm 0.4$ & $39.4 \pm 0.4$ & $31.1 \pm 0.4$ \\
\bottomrule
\end{tabular}

%% file: sections/06-discussions.tex
\section{Discussion and Future Work}\label{sec:discussions}

The conducted analysis of the impact parsers have on ML4SE models shows that it can lead to significant deviations in the models' performance. For TreeLSTM, a model that directly works with ASTs, the difference in F1-score can be as large as 27\% depending on the parser. For Code2Seq, a model that uses path-based representations derived from ASTs, the effect is smaller, but it is still comparable to the metric differences considered to be significant in recent works.

Judging by the results, researchers and practitioners should pay closer attention to parser selection when designing their data processing pipelines. Although some use cases imply strict limitations on the used parser, in most cases there exist several parsers to choose from. In such cases, the choice of a parser should be interpreted as one of hyperparameters when training a model, thus it requires either clear reasoning or empirical validation.

In this work, we focused on the method name prediction task and the Java language. However, the usage of structured code representations proved to be helpful in other tasks as well: variable misuse prediction~\cite{Hellendoorn2020Global}, clone detection~\cite{liang2021ast}, comment generation~\cite{li2020deepcommenter}, etc.
Also, Java is not the only language that has a variety of parsers --- it is also the case for C/C++, Python, PHP, and others. Extension of this study to other tasks and languages is an important topic for future work.

Most structured representations of source code utilize ASTs in some way, but they are not limited to it. Researchers also used data-flow graphs, control-flow graphs, program dependence graphs, and even combined all of them into a single complex structure~\cite{allamanis2017learning, allamanis2020typilus}. As it is the case with parsing, the process of building other structural representations is also non-straightforward. Possible differences in this process range from the exact set of dependencies included in the graph to the specific tools used to extract them. We leave careful analysis of the data preparation process for more complex code representations and its impact on model quality for the future work.

%% file: sections/07-conclusion.tex
\section{Conclusion}\label{sec:conclusion}

With this work, we present an analysis of how the selection of a parsing tool for AST extraction can affect the quality of Machine Learning models that rely on structured representations of code.
We demonstrate that the usage of different parsers leads to the differences in AST characteristics, both in terms of structure (\ie tree size or depth) and content (\ie number of different node types or tokens).
These differences further impact the quality of ML models.
For the TreeLSTM model that works directly with the trees, the difference can reach up to $27\%$ of F1-score and $31\%$ of ChrF for the method name prediction task.
For Code2Seq, a model that uses path-based representation derived from the AST, the difference is less drastic, only about $5\%$ of F1-score and $8\%$ of ChrF, but it is still significant.

In order to simplify experiments with different parsers, we develop \toolname, a tool that can run various parsers in the same manner. In order to run experiments with different parsers, users have to change a single value in the YAML configuration file. The tool facilitates the end-to-end creation of machine learning datasets from source code.
\toolname already supports 5 different languages with 12 parser-language pairs and can be extended to new parsers and languages.
Our tool is publicly available at \url{https://doi.org/10.5281/zenodo.6366591}.

We also publish \javanorm, a dataset similar to the popular \emph{Java-small} one: \url{https://doi.org/10.5281/zenodo.6366599}. Compared to \emph{Java-small}, the validation and testing parts of \javanorm contain diverse projects making the evaluation more robust and representative.
The usage of \javanorm does not require additional computational resources, as it has roughly the same size as \emph{Java-small}, yet it allows more robust comparison between models.

All in all, data preprocessing is one of the critical steps when applying machine learning algorithms.
When working with structural representations of source code (\eg in the form of an AST), data preprocessing becomes rather complex and requires the usage of parsers or other external tools.
To make such studies reproducible while maximizing the achieved quality, it is important to properly select a parsing tool and document the selection process in the reports.
We hope that future researchers will take our findings into consideration and will provide more detailed information about the data preparation process in their works.

%% file: paper.bbl
\begin{thebibliography}{10}
\providecommand{\url}[1]{#1}
\csname url@samestyle\endcsname
\providecommand{\newblock}{\relax}
\providecommand{\bibinfo}[2]{#2}
\providecommand{\BIBentrySTDinterwordspacing}{\spaceskip=0pt\relax}
\providecommand{\BIBentryALTinterwordstretchfactor}{4}
\providecommand{\BIBentryALTinterwordspacing}{\spaceskip=\fontdimen2\font plus
\BIBentryALTinterwordstretchfactor\fontdimen3\font minus
  \fontdimen4\font\relax}
\providecommand{\BIBforeignlanguage}[2]{{%
\expandafter\ifx\csname l@#1\endcsname\relax
\typeout{** WARNING: IEEEtran.bst: No hyphenation pattern has been}%
\typeout{** loaded for the language `#1'. Using the pattern for}%
\typeout{** the default language instead.}%
\else
\language=\csname l@#1\endcsname
\fi
#2}}
\providecommand{\BIBdecl}{\relax}
\BIBdecl

\bibitem{sharma2021survey}
T.~Sharma, M.~Kechagia, S.~Georgiou, R.~Tiwari, and F.~Sarro, ``A survey on
  machine learning techniques for source code analysis,'' \emph{arXiv preprint
  arXiv:2110.09610}, 2021.

\bibitem{Ernst2017nlp_for_se}
M.~D. Ernst, ``Natural language is a programming language: Applying natural
  language processing to software development,'' in \emph{SNAPL 2017: the 2nd
  Summit oN Advances in Programming Languages}, Asilomar, CA, USA, May 2017,
  pp. 4:1--4:14.

\bibitem{alon2018code2seq}
U.~Alon, S.~Brody, O.~Levy, and E.~Yahav, ``code2seq: Generating sequences from
  structured representations of code,'' \emph{arXiv preprint arXiv:1808.01400},
  2018.

\bibitem{allamanis2020typilus}
M.~Allamanis, E.~T. Barr, S.~Ducousso, and Z.~Gao, ``Typilus: Neural type
  hints,'' in \emph{PLDI}, 2020.

\bibitem{zugner2021codetrans}
D.~Z{\"u}gner, T.~Kirschstein, M.~Catasta, J.~Leskovec, and S.~G{\"u}nnemann,
  ``Language-agnostic representation learning of source code from structure and
  context,'' in \emph{International Conference on Learning Representations
  (ICLR)}, 2021.

\bibitem{alon2018general}
U.~Alon, M.~Zilberstein, O.~Levy, and E.~Yahav, ``A general path-based
  representation for predicting program properties,'' \emph{ACM SIGPLAN
  Notices}, vol.~53, no.~4, pp. 404--419, 2018.

\bibitem{allamanis2017learning}
M.~Allamanis, M.~Brockschmidt, and M.~Khademi, ``Learning to represent programs
  with graphs,'' \emph{arXiv preprint arXiv:1711.00740}, 2017.

\bibitem{Hellendoorn2020Global}
\BIBentryALTinterwordspacing
V.~J. Hellendoorn, C.~Sutton, R.~Singh, P.~Maniatis, and D.~Bieber, ``Global
  relational models of source code,'' in \emph{International Conference on
  Learning Representations}, 2020. [Online]. Available:
  \url{https://openreview.net/forum?id=B1lnbRNtwr}
\BIBentrySTDinterwordspacing

\bibitem{rusak2018ast}
G.~Rusak, A.~Al-Dujaili, and U.-M. O'Reilly, ``Ast-based deep learning for
  detecting malicious powershell,'' in \emph{Proceedings of the 2018 ACM SIGSAC
  Conference on Computer and Communications Security}, 2018, pp. 2276--2278.

\bibitem{max_brunsfeld_2022_6326492}
\BIBentryALTinterwordspacing
M.~Brunsfeld, P.~Thomson, A.~Hlynskyi, J.~Vera, P.~Turnbull, T.~Clem,
  D.~Creager, A.~Helwer, R.~Rix, H.~van Antwerpen, M.~Davis, Ika, T.-A. Nguyen,
  S.~Brunk, N.~Hasabnis, bfredl, M.~Dong, V.~Panteleev, ikrima, S.~Kalt,
  K.~Lampe, A.~Pinkus, M.~Schmitz, M.~Krupcale, narpfel, S.~Gallegos,
  V.~Martí, Edgar, and G.~Fraser, ``tree-sitter/tree-sitter: v0.20.6,'' Mar.
  2022. [Online]. Available: \url{https://doi.org/10.5281/zenodo.6326492}
\BIBentrySTDinterwordspacing

\bibitem{10.5555/2501720}
T.~Parr, \emph{The Definitive ANTLR 4 Reference}, 2nd~ed.\hskip 1em plus 0.5em
  minus 0.4em\relax Pragmatic Bookshelf, 2013.

\bibitem{DBLP:journals/corr/TaiSM15}
\BIBentryALTinterwordspacing
K.~S. Tai, R.~Socher, and C.~D. Manning, ``Improved semantic representations
  from tree-structured long short-term memory networks,'' \emph{CoRR}, vol.
  abs/1503.00075, 2015. [Online]. Available:
  \url{http://arxiv.org/abs/1503.00075}
\BIBentrySTDinterwordspacing

\bibitem{hochreiter1997lstm}
\BIBentryALTinterwordspacing
S.~Hochreiter and J.~Schmidhuber, ``Long short-term memory,'' \emph{Neural
  Comput.}, vol.~9, no.~8, p. 1735–1780, nov 1997. [Online]. Available:
  \url{https://doi.org/10.1162/neco.1997.9.8.1735}
\BIBentrySTDinterwordspacing

\bibitem{nagar2021code}
A.~R. Nagar, \emph{Code Search Using Code2Seq}.\hskip 1em plus 0.5em minus
  0.4em\relax University of California, Irvine, 2021.

\bibitem{Zhang2020CBPath2Vec}
X.~Zhang, Y.~Lu, and K.~Shi, ``Cb-path2vec: A cross block path based
  representation for software defect prediction,'' \emph{2020 IEEE 6th
  International Conference on Computer and Communications (ICCC)}, pp.
  1961--1966, 2020.

\bibitem{fernandes2018structured}
\BIBentryALTinterwordspacing
P.~Fernandes, M.~Allamanis, and M.~Brockschmidt, ``Structured neural
  summarization,'' in \emph{International Conference on Learning
  Representations}, 2019. [Online]. Available:
  \url{https://openreview.net/forum?id=H1ersoRqtm}
\BIBentrySTDinterwordspacing

\bibitem{kovalenko2019pathminer}
V.~Kovalenko, E.~Bogomolov, T.~Bryksin, and A.~Bacchelli, ``Pathminer: a
  library for mining of path-based representations of code,'' in
  \emph{Proceedings of the 16th International Conference on Mining Software
  Repositories}.\hskip 1em plus 0.5em minus 0.4em\relax IEEE Press, 2019, pp.
  13--17.

\bibitem{psi}
``Program structure interface,''
  \url{https://jetbrains.org/intellij/sdk/docs/basics/architectural_overview/psi.html},
  [Online; accessed 17-March-2022].

\bibitem{jdt}
J.~Arthanareeswaran, K.~P. Tatavarthi, M.~Palat, N.~Gupta, and S.~Sinha,
  ``{Eclipse Java Development tools},'' \url{https://www.eclipse.org/jdt/},
  [Online; accessed 17-March-2022].

\bibitem{danny_van_bruggen_2020_3842713}
\BIBentryALTinterwordspacing
D.~van Bruggen, F.~Tomassetti, R.~Howell, M.~Langkabel, N.~Smith, A.~Bosch,
  M.~Skoruppa, C.~Maximilien, ThLeu, Panayiotis, S.~K. (@skirsch79), Simon,
  J.~Beleites, W.~Tibackx, jean~pierre L, A.~Rouél, edefazio, D.~Schipper,
  Mathiponds, W.~you want~to know, R.~Beckett, ptitjes, kotari4u, M.~Wyrich,
  R.~Morais, M.~Coene, bresai, Implex1v, and B.~Haumacher,
  ``{javaparser/javaparser: Release javaparser- parent-3.16.1},'' May 2020.
  [Online]. Available: \url{https://doi.org/10.5281/zenodo.3842713}
\BIBentrySTDinterwordspacing

\bibitem{chen2018tree2tree}
X.~Chen, C.~Liu, and D.~Song, ``Tree-to-tree neural networks for program
  translation,'' ser. NIPS'18.\hskip 1em plus 0.5em minus 0.4em\relax Red Hook,
  NY, USA: Curran Associates Inc., 2018, p. 2552–2562.

\bibitem{shido2019automatic}
Y.~Shido, Y.~Kobayashi, A.~Yamamoto, A.~Miyamoto, and T.~Matsumura, ``Automatic
  source code summarization with extended tree-lstm,'' in \emph{2019
  International Joint Conference on Neural Networks (IJCNN)}.\hskip 1em plus
  0.5em minus 0.4em\relax IEEE, 2019, pp. 1--8.

\bibitem{liang2021ast}
H.~Liang and L.~Ai, ``Ast-path based compare-aggregate network for code clone
  detection,'' in \emph{2021 International Joint Conference on Neural Networks
  (IJCNN)}.\hskip 1em plus 0.5em minus 0.4em\relax IEEE, 2021, pp. 1--8.

\bibitem{shi2020pathpair2vec}
K.~Shi, Y.~Lu, J.~Chang, and Z.~Wei, ``Pathpair2vec: An ast path pair-based
  code representation method for defect prediction,'' \emph{Journal of Computer
  Languages}, vol.~59, p. 100979, 2020.

\bibitem{alon2019code2vec}
U.~Alon, M.~Zilberstein, O.~Levy, and E.~Yahav, ``code2vec: Learning
  distributed representations of code,'' \emph{Proceedings of the ACM on
  Programming Languages}, vol.~3, no. POPL, pp. 1--29, 2019.

\bibitem{Zhang2020retrieval_cs}
\BIBentryALTinterwordspacing
J.~Zhang, X.~Wang, H.~Zhang, H.~Sun, and X.~Liu, ``Retrieval-based neural
  source code summarization,'' in \emph{Proceedings of the ACM/IEEE 42nd
  International Conference on Software Engineering}, ser. ICSE '20.\hskip 1em
  plus 0.5em minus 0.4em\relax New York, NY, USA: Association for Computing
  Machinery, 2020, p. 1385–1397. [Online]. Available:
  \url{https://doi.org/10.1145/3377811.3380383}
\BIBentrySTDinterwordspacing

\bibitem{mou2016tbcnn}
L.~Mou, G.~Li, L.~Zhang, T.~Wang, and Z.~Jin, ``Convolutional neural networks
  over tree structures for programming language processing,'' in
  \emph{Proceedings of the Thirtieth AAAI Conference on Artificial
  Intelligence}, ser. AAAI'16.\hskip 1em plus 0.5em minus 0.4em\relax AAAI
  Press, 2016, p. 1287–1293.

\bibitem{vaswani2017transformer}
A.~Vaswani, N.~Shazeer, N.~Parmar, J.~Uszkoreit, L.~Jones, A.~N. Gomez,
  L.~Kaiser, and I.~Polosukhin, ``Attention is all you need,'' in
  \emph{Proceedings of the 31st International Conference on Neural Information
  Processing Systems}, ser. NIPS'17.\hskip 1em plus 0.5em minus 0.4em\relax Red
  Hook, NY, USA: Curran Associates Inc., 2017, p. 6000–6010.

\bibitem{DBLP:conf/kbse/FalleriMBMM14}
\BIBentryALTinterwordspacing
J.~Falleri, F.~Morandat, X.~Blanc, M.~Martinez, and M.~Monperrus,
  ``Fine-grained and accurate source code differencing,'' in \emph{{ACM/IEEE}
  International Conference on Automated Software Engineering, {ASE} '14,
  Vasteras, Sweden - September 15 - 19, 2014}, 2014, pp. 313--324. [Online].
  Available: \url{http://doi.acm.org/10.1145/2642937.2642982}
\BIBentrySTDinterwordspacing

\bibitem{yin2017syntactic}
P.~Yin and G.~Neubig, ``A syntactic neural model for general-purpose code
  generation,'' \emph{arXiv preprint arXiv:1704.01696}, 2017.

\bibitem{wei2019cg-cs}
\BIBentryALTinterwordspacing
B.~Wei, G.~Li, X.~Xia, Z.~Fu, and Z.~Jin, ``\BIBforeignlanguage{English}{Code
  generation as a dual task of code summarization},'' in
  \emph{\BIBforeignlanguage{English}{NIPS Proceedings - Advances in Neural
  Information Processing Systems 32 (NIPS 2019)}}, ser. Advances in Neural
  Information Processing Systems, H.~Wallach, H.~Larochelle, A.~Beygelzimer,
  F.~d'AlcheBuc, E.~Fox, and R.~Garnett, Eds., vol.~32.\hskip 1em plus 0.5em
  minus 0.4em\relax Neural Information Processing Systems (NIPS), 2019,
  advances in Neural Information Processing Systems 2019, NIPS 2019 ;
  Conference date: 08-12-2019 Through 14-12-2019. [Online]. Available:
  \url{https://nips.cc/Conferences/2019,
  https://papers.nips.cc/book/advances-in-neural-information-processing-systems-32-2019}
\BIBentrySTDinterwordspacing

\bibitem{tairas2006phoenix}
R.~Tairas and J.~Gray, ``Phoenix-based clone detection using suffix trees,'' in
  \emph{Proceedings of the 44th annual Southeast regional conference}, 2006,
  pp. 679--684.

\bibitem{tunes}
C.~Tunes, ``javalang: Pure python java parser and tools,''
  \url{https://github.com/c2nes/javalang}, [Online; accessed 17-March-2022].

\bibitem{psiminer}
E.~Spirin, E.~Bogomolov, V.~Kovalenko, and T.~Bryksin, ``Psiminer: A tool for
  mining rich abstract syntax trees from code,'' 05 2021, pp. 13--17.

\bibitem{pawlak:hal-01169705}
\BIBentryALTinterwordspacing
R.~Pawlak, M.~Monperrus, N.~Petitprez, C.~Noguera, and L.~Seinturier, ``{Spoon:
  A Library for Implementing Analyses and Transformations of Java Source
  Code},'' \emph{{Software: Practice and Experience}}, vol.~46, pp. 1155--1179,
  2015. [Online]. Available:
  \url{https://hal.archives-ouvertes.fr/hal-01078532/document}
\BIBentrySTDinterwordspacing

\bibitem{xuan2016nopol}
J.~Xuan, M.~Martinez, F.~Demarco, M.~Clement, S.~L. Marcote, T.~Durieux,
  D.~Le~Berre, and M.~Monperrus, ``Nopol: Automatic repair of conditional
  statement bugs in java programs,'' \emph{IEEE Transactions on Software
  Engineering}, vol.~43, no.~1, pp. 34--55, 2016.

\bibitem{martinez2016astor}
M.~Martinez and M.~Monperrus, ``Astor: A program repair library for java,'' in
  \emph{Proceedings of the 25th International Symposium on Software Testing and
  Analysis}, 2016, pp. 441--444.

\bibitem{long2017automatic}
F.~Long, P.~Amidon, and M.~Rinard, ``Automatic inference of code transforms for
  patch generation,'' in \emph{Proceedings of the 2017 11th Joint Meeting on
  Foundations of Software Engineering}, 2017, pp. 727--739.

\bibitem{collard2013srcml}
M.~L. Collard, M.~J. Decker, and J.~I. Maletic, ``srcml: An infrastructure for
  the exploration, analysis, and manipulation of source code: A tool
  demonstration,'' in \emph{2013 IEEE International Conference on Software
  Maintenance}.\hskip 1em plus 0.5em minus 0.4em\relax IEEE, 2013, pp.
  516--519.

\bibitem{allamanis2015acc_method_names}
\BIBentryALTinterwordspacing
M.~Allamanis, E.~T. Barr, C.~Bird, and C.~Sutton, ``Suggesting accurate method
  and class names,'' in \emph{Proceedings of the 2015 10th Joint Meeting on
  Foundations of Software Engineering}, ser. ESEC/FSE 2015.\hskip 1em plus
  0.5em minus 0.4em\relax New York, NY, USA: Association for Computing
  Machinery, 2015, p. 38–49. [Online]. Available:
  \url{https://doi.org/10.1145/2786805.2786849}
\BIBentrySTDinterwordspacing

\bibitem{student1908probable}
Student, ``The probable error of a mean,'' \emph{Biometrika}, pp. 1--25, 1908.

\bibitem{roy2021reassessing}
D.~Roy, S.~Fakhoury, and V.~Arnaoudova, ``Reassessing automatic evaluation
  metrics for code summarization tasks,'' in \emph{Proceedings of the 29th ACM
  Joint Meeting on European Software Engineering Conference and Symposium on
  the Foundations of Software Engineering}, 2021, pp. 1105--1116.

\bibitem{banerjee2005meteor}
S.~Banerjee and A.~Lavie, ``Meteor: An automatic metric for mt evaluation with
  improved correlation with human judgments,'' in \emph{Proceedings of the acl
  workshop on intrinsic and extrinsic evaluation measures for machine
  translation and/or summarization}, 2005, pp. 65--72.

\bibitem{papineni2002bleu}
K.~Papineni, S.~Roukos, T.~Ward, and W.-J. Zhu, ``Bleu: a method for automatic
  evaluation of machine translation,'' in \emph{Proceedings of the 40th annual
  meeting of the Association for Computational Linguistics}, 2002, pp.
  311--318.

\bibitem{lin2004looking}
C.-Y. Lin and F.~Och, ``Looking for a few good metrics: Rouge and its
  evaluation,'' in \emph{Ntcir workshop}, 2004.

\bibitem{popovic2015chrf}
M.~Popovi{\'c}, ``chrf: character n-gram f-score for automatic mt evaluation,''
  in \emph{Proceedings of the Tenth Workshop on Statistical Machine
  Translation}, 2015, pp. 392--395.

\bibitem{zhang2019bertscore}
T.~Zhang, V.~Kishore, F.~Wu, K.~Q. Weinberger, and Y.~Artzi, ``Bertscore:
  Evaluating text generation with bert,'' \emph{arXiv preprint
  arXiv:1904.09675}, 2019.

\bibitem{Efron93bootstrap}
B.~Efron and R.~J. Tibshirani, \emph{An Introduction to the Bootstrap}, ser.
  Monographs on Statistics and Applied Probability.\hskip 1em plus 0.5em minus
  0.4em\relax Boca Raton, Florida, USA: Chapman \& Hall/CRC, 1993, no.~57.

\bibitem{feng2020codebert}
\BIBentryALTinterwordspacing
Z.~Feng, D.~Guo, D.~Tang, N.~Duan, X.~Feng, M.~Gong, L.~Shou, B.~Qin, T.~Liu,
  D.~Jiang, and M.~Zhou, ``{C}ode{BERT}: A pre-trained model for programming
  and natural languages,'' in \emph{Findings of the Association for
  Computational Linguistics: EMNLP 2020}.\hskip 1em plus 0.5em minus
  0.4em\relax Online: Association for Computational Linguistics, Nov. 2020, pp.
  1536--1547. [Online]. Available:
  \url{https://aclanthology.org/2020.findings-emnlp.139}
\BIBentrySTDinterwordspacing

\bibitem{li2020deepcommenter}
B.~Li, M.~Yan, X.~Xia, X.~Hu, G.~Li, and D.~Lo, ``Deepcommenter: a deep code
  comment generation tool with hybrid lexical and syntactical information,'' in
  \emph{Proceedings of the 28th ACM Joint Meeting on European Software
  Engineering Conference and Symposium on the Foundations of Software
  Engineering}, 2020, pp. 1571--1575.

\end{thebibliography}
